# Understanding the Twitter Usage of Science Citation Index (SCI) Journals

Aravind Sesagiri Raamkumar[0000-0001-7200-7787], Mojisola Erdt[0000-0003-2371-6768], Harsha Vijayakumar, Aarthy Nagarajan, and Yin-Leng Theng [0000-0003-2351-8884]

Wee Kim Wee School of Communication and Information,
Nanyang Technological University, Singapore 637718, Singapore
{aravind0002, mojisola.erdt, hvijayakumar, aarthyn, tyltheng}@ntu.edu.sg

**Abstract.** This paper investigates the Twitter interaction patterns of journals from the Science Citation Index (SCI) of Master Journal List (MJL). A total of 953,253 tweets extracted from 857 journal accounts, were analyzed in this study. Findings indicate that SCI journals interacted more with each other but much less with journals from other citation indices. The network structure of the communication graph resembled a tight crowd network, with Nature journals playing a major part. Information sources such as news portals and scientific organizations were mentioned more in tweets, than academic journal Twitter accounts. Journals with high journal impact factors (JIFs) were found to be prominent hubs in the communication graph. Differences were found between the Twitter usage of SCI journals with Humanities and Social Sciences (HSS) journals.

## 1 Introduction

The scientific community has been interested in Twitter due to the scope for extended information dissemination and promotion of research [1]. To enable discussions about research papers, academic journals use Twitter [2–4]. Furthermore, journals with social media accounts (e.g., Twitter) have been found to have higher research metrics [5]. The current exploratory study attempts to compliment two earlier Twitter studies conducted with tweets posted by academic journals. In the first study [6], the overall social media presence of academic journals was investigated, specifically for Twitter and Facebook platforms. In the second study [7], the Twitter usage of Arts & Humanities Citation Index (AHCI) and Social Science Citation Index (SSCI) journals was reported. This study reports the Twitter interaction dynamics of Science Citation Index (SCI) academic journals indexed in the Master Journal List (MJL). In the MJL, the journals are indexed in three main citation indices namely SCI, AHCI and SSCI. The following research questions are investigated in this study.

*RQ1: Does the communication between journals in Twitter conversations of SCI journals differ from HSS (AHCI & SSCI) journals?*

*RQ2: Which network structure best represents the SCI journals Twitter communication graph?*

*RQ3: What type of Twitter accounts are top authorities in the communication graph of SCI journals?*

*RQ4: What disciplines do the top SCI journals in the communication graph represent and does their Twitter popularity reflect their ranking in the Journal Citation Report (JCR)?*

## 2   Methodology

### 2.1   Base Dataset

A subset of the dataset used in a previous study [6] was used for this current study. The dataset preparation is described as follows. Journal titles from SCI were extracted from the MJL. After manually identifying the Twitter accounts of the journals, the tweets were extracted using the Twitter API. The basic Twitter API permitted extraction of only 3,000 tweets for each Twitter account during the time of data collection. Out of the 8,827 SCI journals in MJL, 857 were found to have Twitter accounts and 953,253 tweets were extracted for these journals.

### 2.2   Data Preparation

The data for addressing the four research questions was prepared through the following process. In this paper's context, the term 'mention' refers to a Twitter account tagged in a tweet. The tweets that contained mentions were shortlisted for the next step. From these tweets, the parent Twitter account name (journal) and the mention(s) were extracted. A combination of Twitter account name and the mention is referred to as a 'conversation' in this paper. It is to be noted that a single tweet could contain multiple mentions hence there could be multiple conversations within a tweet. Using the conversations data from the previous step, a Twitter communication graph was built. Gephi [8] was used to build this graph. In the communication graph, the source node is always the journal Twitter account whereas the target node is the mention (a Twitter account which is not necessarily a journal). The nodes with a degree value of one were filtered out, for handling the issue of sparsity. After building the graph, the community detection algorithm ForeAtlas2 was used in Gephi to identify the different communities in the graph. In this algorithm, communities are identified based on the frequency of interactions between the source and target nodes of the graph. A sub-graph of a particular set of nodes that interact more with each other than with other nodes, is regarded collectively as a community. The top 20 nodes (Twitter accounts) with the highest in-degree values and out-degree values in the communication graph were identified as a part of this step. The in-degree of a graph is the number of edges where the node is a target node while the out-degree of a node is the number of edges where the node is the source node. A node with a high in-degree value is referred to as an 'authority' while a node with a high out-degree value is referred to as a 'hub'. The nodes with the highest out-degrees (hubs) are the Twitter accounts managed by journals whereas the nodes with the highest in-degrees (authorities) could be any Twitter account.

## 3 Results

### 3.1 Twitter Conversations

SCI journals' conversation stats are listed in Table 1. The number of conversations in SCI journals was high ($n$=509,062) corresponding to the high number of tweets (refer Section 2.1). This metric was nearly five times more than AHCI ($n$=103,181) and SSCI ($n$= 121,099) journals. To determine the interaction levels with other journals, we identified the mentions which were journal accounts (data listed in rows D, E, F of Table 1). SCI journals had the highest percentage of conversations at an intra-index level (18.16%). In [7], it was identified that HSS journals intra-index conversations had comparatively lower percentages (13.39% for AHCI, 13.25% for SSCI).

Table 1. SCI journals' Twitter conversation statistics

| Entity | $n$ |
|---|---|
| Conversations [A] | 509,062 |
| Unique journal accounts [B] | 768 |
| Unique user-mentions [C] | 88,021 |
| Conversations where user-mentions are SCI journals [D] | 92,421 (18.16% of A) |
| Conversations where user-mentions are AHCI journals [E] | 88 |
| Conversations where user-mentions are SSCI journals [F] | 592 |

### 3.2 Communication Graph

The communication graph[1] generated for the SCI journals is illustrated in Figure 2. The node color represents the parent community of a node. The size of the nodes was set according to the betweenness centrality values of the nodes. Representative labels have been added for each community based on the commonality of topics[2]. Communities comprised of journal accounts belonging mainly to the disciplines of Chemistry, Bioscience, Ecology and Surgery. Technology & Engineering formed a very small community in this graph. Interestingly, there was one community which was entirely oriented towards journals from the Nature Publishing Group (NPG). The SCI graph structure resembles a "tight crowd" [9] mainly because of the central presence of Nature journals across many communities. This type of network has been observed in other

---

[1] Higher resolution image can be viewed at this link https://bit.ly/2KPo8EH
[2] The topics were identified based on the profile description in the Twitter accounts

related Twitter scholarly communication studies [10]. Unlike the SCI graph, the communities in the AHCI and SSCI graphs were represented by a greater number of topics [7]. Both these graphs were classified as a "community clusters" due to the presence of multiple communities and minimal interspersed nodes.

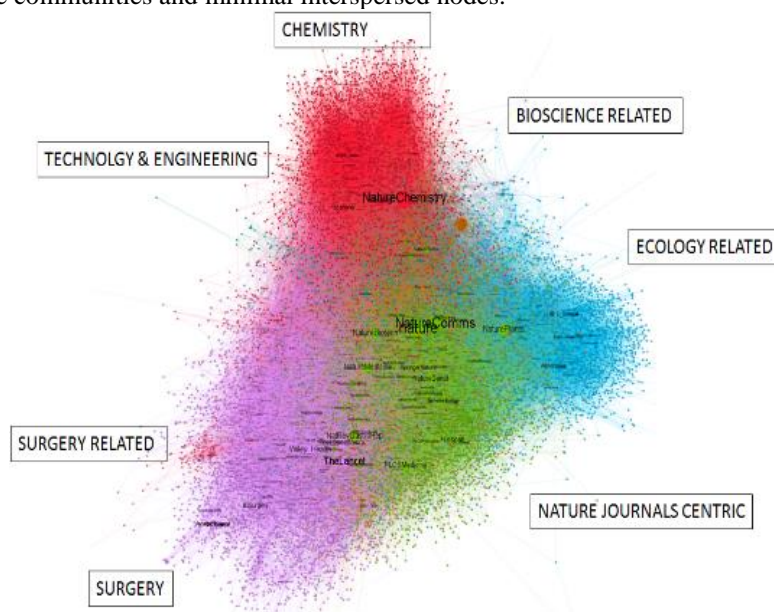

**Figure 1.** SCI Twitter communication graph.

### 3.3 Top Authorities and Hubs

In Table 2, the top 20 authorities (based on in-degree values) from the SCI communication graph are listed, along with the account types which had been identified for these Twitter accounts. The major presence of news portals as top authorities was noticed (e.g., Nature News, The Guardian), with these accounts occupying 25% ($n=5$) of the top 20 authorities. Scientific organization was another authority type identified, which was as popular ($n=5$) as news portals. We used this account type as an umbrella term covering different organizations related to academia and scientific research. The next most popular authority type identified was journal. SCI had the most number of journals ($n=4$), whereas AHCI and SSCI had only one and two journals in the top 20 authorities list respectively in the earlier study [7]. Agencies were the next most popular authority type ($n=3$) identified. These Twitter accounts included government and global agencies (WHO, NIH & CDC). Magazine Twitter accounts ($n=2$) were not as popular as authorities for SCI journals, unlike HSS journals where this account type was quite prevalent (approximately 35% of the top 20) [7].

**Table 2.** Top 20 authorities in the SCI communication graph

| Twitter Handle | Account Name | Account Type | in* |
|---|---|---|---|
| NatureNews | Nature News & Comment | News | 209 |
| guardian | The Guardian | News | 172 |
| sciencemagazine | Science Magazine | Journal | 165 |
| nytimes | The New York Times | News | 159 |
| sciam | Scientific American | Magazine | 157 |
| WHO | World Health Organization | Specialized Agency | 142 |
| newscientist | New Scientist | Magazine | 130 |
| nature | Nature | Journal | 127 |
| altmetric | Altmetric | Scientific Organization | 123 |
| NIH | NIH | Government Agency | 122 |
| YouTube | YouTube | Video Sharing | 114 |
| wellcometrust | Wellcome Trust | Scientific Organization | 112 |
| MayoClinic | Mayo Clinic | Scientific Organization | 109 |
| NEJM | NEJM | Journal | 104 |
| physorg_com | Phys.org | News | 102 |
| UniofOxford | Oxford University | Scientific Organization | 102 |
| guardianscience | Guardian Science | News | 102 |
| royalsociety | The Royal Society | Scientific Organization | 101 |
| CDCgov | CDC | Government Agency | 100 |
| TheLancet | The Lancet | Journal | 96 |

*Note: in\* refers to in-degree value*

In Table 3, the top 20 hubs (based on out-degree values) from the SCI communication graph are listed. It is well-known that the JIF is measured and indexed along with the corresponding quartiles in the JCR [11] every year. We have included the JIF quartile data for the top 20 hubs in Table 4. The JIF quartile indicates the popularity of the journal within the citation network. For instance, if a journal belongs to the first quartile, it is considered to be among the top 25 percentile of journals for the respective research subject. Journals from the disciplines of medicine ($n=3$), biology ($n=4$) and chemistry ($n=4$) were prominent. It can be observed that most of the journals ($n=14$) in this list were from the first quartile of their respective research subjects. This finding indicates that the top journals in SCI seem to be most active in Twitter. Interestingly, engineering journal accounts were not present in this list.

**Table 3.** Top 20 hubs in the SCI communication graph

| Twitter Handle | Journal Name | out* | JQ^ |
|---|---|---|---|
| BiolJLinnSoc | Biological Journal of the Linnean Society | 835 | 3 |
| BMCMedicine | BMC Medicine | 769 | 1 |
| ChemMater | Chemistry of Materials | 731 | 1 |
| acsnano | ACS Nano | 716 | 1 |
| BiochemJ | Biochemical Journal | 714 | 2 |
| clin_sci | Clinical Science | 682 | 1 |
| ChemCatChem | Chemcatchem | 661 | 1 |
| PLOSPathogens | PLOS Pathogens | 642 | 1 |
| NatureChemistry | Nature Chemistry | 636 | 1 |
| FunEcology | Functional Ecology | 626 | 1 |
| PLOSMedicine | PLOS Medicine | 605 | 1 |
| NaturePlants | Nature Plants | 603 | 4 |
| JExpMed | Journal of Experimental Medicine | 576 | 1 |
| eLife | eLife | 574 | 1 |
| ASIHCopeia | Copeia | 543 | 2 |
| NatureGenet | Nature Genetics | 526 | 1 |
| GenomeBiology | Genome Biology | 510 | 1 |
| BiosciReports | Bioscience Reports | 506 | 3 |
| cenmag | Chemical & Engineering News | 485 | 4 |
| TheLancetInfDis | Lancet Infectious Diseases | 485 | 1 |

*Note: out\* refers to out-degree value; JQ^ refers to JIF quartile; OA refers to fully Open Access journals*

## 4  Conclusion

In the current study's context, SCI journals were found to have a marginally better presence than HSS journals on Twitter. For RQ1, the conversation statistics revealed certain differences between the journals from SCI and HSS. In our earlier study [7], the structure of the communication graphs of both AHCI and SSCI journals largely resembled a 'community clusters' network where there were multiple communities with high intercommunication. In the current study, the bigger SCI graph resembled a 'tight crowd' network. We can conclude that there are differences at the conversational level in Twitter between hard (SCI) and non-hard (AHCI & SSCI) sciences for RQ1 and RQ2.

The top authorities and hubs in the communication graph were identified for RQ3 and RQ4. News portals and scientific organizations were amongst the most authoritative Twitter accounts. Magazines such as the New Scientist and Scientific American were also present in this authorities list. Compared to HSS journals, SCI had more number of journals as authoritative accounts. With the participation of different account types such as news portals, scientific organizations, journals, magazines, and government agencies, the involvement of key players in the scholarly communication lifecycle of SCI journals is clearly perceived. This is in contrast to the HSS journals' top authorities lists which comprised mostly of news portals and magazines which cater to the general public [7]. Interestingly, three Twitter accounts were found to be present in the top 20 authority lists of both SCI and HSS journals. These accounts were Guardian, NY Times and YouTube. The Top 20 Twitter hubs list comprised almost entirely of natural sciences and medicine related journals. This finding highlights the prominent outreach activities of journals from these disciplines on Twitter. With regards to the question of ascertaining whether the top hubs in Twitter were also top journals in the citation network (represented by JCR rankings), data shows that SCI had mostly first quartile journals in the list, unlike SSCI journals where journals from all four quartiles were present in the list [7].

## 5  Acknowledgements

The research project "Altmetrics: Rethinking And Exploring New Ways Of Measuring Research" is supported by the National Research Foundation, Prime Minister's Office, Singapore under its Science of Research, Innovation and Enterprise programme (SRIE Award No. NRF2014-NRF-SRIE001-019).